\documentstyle[aps,twocolumn,epsf]{revtex}

\begin{document}
\title{The Phase Diagrams of F=2 Spinor Bose Condensates} 
\author{C.V. Ciobanu $^1$, S.-K. Yip$^{2,1}$ and Tin-Lun Ho$^1$}
\address{$^1$Department of Physics,  The Ohio State University,
Columbus, Ohio 43210\newline $^2$Physics Division, National 
Centre for Theoretical Sciences, P.O. Box 2-131, Hinschu, Taiwan 300}
\maketitle

\begin{abstract}
We show that there are three possible phases for a spin-2 
spinor Bose condensate, one more compared to the spin-1 case. 
The order parameters of these phases are the spontaneous magnetization and 
the singlet pair amplitude. Current estimates of scattering lengths 
show that all three phases have realizations in optically trapped 
alkali atoms. There is also a 
one-to-one correspondence between the structure of a spin-2 spinor Bose 
condensate and that of a d-wave BCS superfluid.
\end{abstract}

One of the recent major developments in Bose-Einstein Condensation 
(BEC) in atomic gases is the study of 
dilute Bose gases with internal degrees of freedom. The first realization of
such system is found in optically trapped $^{23}$Na, which is a spin-1 Bose
gas\cite{MIT}.  Recently, JILA has also created a ``spin"-1/2 
Bose gas by continually cycling between the $F=1$ and $F=2$ states of 
magnetically trapped $^{87}$Rb\cite{spin1/2}. 
In the case of spin-1 Bose gas, the nature of the 
spinor condensate depends crucially on the magnetic interaction. In zero
magnetic field, the spinor condensate can be either ferromagnetic or 
``polar", which has very different properties\cite{Hospinor}\cite{Jap}. 

Generally, only atoms in the low lying hyperfine multiplet are confined in the 
optical trap. Those in the higher hyperfine multiplet will leave 
the trap by spin-flip scattering. In case of $^{23}$Na and $^{87}$Rb, 
their hyperfine multiplets ($F=2$ and $F=1$) are regular, i.e. the 
higher spin state ($F=2)$ has higher energy. 
Since spin-flip scattering is strong in $^{23}$Na, it may be difficult to 
produce a spin-2 Bose gas in this system. 
On the other hand, $^{87}$Rb has much weaker spin-flip scattering and is 
a candidate for optically trapped spin-2 Bose
gas. In the case of $^{85}$Rb, the lower multiplet has spin $F=2$. 
It also has a negative
s-wave scattering length in zero field.  Should the current effort
to Bose condense $^{85}$Rb in magnetic traps be successful, it is
conceivable that an $F=2$ spinor condensate can be trapped optically in 
low fields, if the three particle losses when the field is reduced through 
the Feshbach resonance is not too large. 

In this paper, we study the ground state structure of a spin-2 Bose
gas within the single condensate approximation. 
In the case of spin-1 Bose gas, it has been realized recently that the ground
state can be ``fragmented", (i.e 
containing more than one condensate)\cite{Hofrag}\cite{Bigelow}. 
Despite this fact, the phase diagram for single spinor condensates
remains highly valuable and in fact gives the best agreement with 
experiments so far\cite{MIT}. This is because the spin-1 fragmented state is 
delicate with respect to spin-nonconserving perturbations, which will drive 
the system toward a single condensate state.  For these reasons, we shall 
first focus on the ground states of single spinor condensates. 
We shall consider only linear Zeeman effect, which is already much 
more subtle than the spin-1 case.
The actual fragmented structures as well as 
quadratic Zeeman effects will be discussed elsewhere. 

Because of the increase in spin value, spin-2 Bose gas has one more 
interaction parameter than that of spin-1 Bose gas. As a result, 
there are three possible phases in zero magnetic field 
(instead of two in the spin-1 case). These phases are characterized by 
a pair of order parameters $\langle {\bf f} \rangle$ and
$\Theta$ describing the ferromagnetic order and the formation of singlet pairs
respectively. The order parameter $\Theta$ is absent in the spin-1 case. 
The first two phases, which are characterised by 
($|\langle {\bf f} \rangle| = 2, 
\Theta =0$) and ($\langle {\bf f} \rangle=0, |\Theta| = 1$) in zero field,  are 
referred to as ferromagnetic and polar phases, respectively. They 
are the analogs of the corresponding phases in the spin-1 case. 
The third phase, ($\langle {\bf f} \rangle=0, \Theta=0$) 
is a non-magnetic but degenerate set of states which has no spin-1 analog. 
They will be referred to as the 
``cyclic" states because of their close analog to the d-wave BCS
superfluids which we shall discuss at the end. In finite fields (along $z$), 
both $\langle f_{z} \rangle$ and $\Theta$ are non-zero for all three
phases. From the current estimates of 
scattering lengths,  we find that the spin-2 Bosons of $^{87}$Rb and 
$^{23}$Na will have a ``polar" ground state, whereas those of 
$^{85}$Rb and $^{83}$Rb will be cyclic and ferromagnetic respectively. 
All these phases can be distinguished by their very different 
density profiles. \cite{March}

{\em Low-energy Hamiltonian}. The effective low-energy
Hamiltonian of a spin-$f$ Bose gas was derived earlier\cite{Hospinor} with 
particle interactions of the form 
${\cal V}({\bf r}_{1}-{\bf r}_{2})$$=$$\delta( {\bf r}_{1}-{\bf r}_{2}) 
\sum_{F=0}^{2f} g_{F}{\cal P}_{F}$, $g_{F}= 4\pi\hbar^2 a_{F}/M$, where
$M$ is the mass of the atom, ${\cal P}_{F}$ is the projection operator 
which projects the pair 1 and 2 into a total hyperfine spin $F$ state, 
and $a_{F}$ the s-wave scattering length in the total spin $F$ channel. 
Symmetry implies that only even $F$ terms appear in ${\cal V}$. 
For spin $f=2$ Bosons, we have 
${\cal V} = g_4{\cal P}_4+g_{2}{\cal P}_{2} + g_{0}{\cal P}_{0}$. 
Using the fact that ${\cal P}_0+{\cal P}_2+{\cal P}_4=1$ \cite{Hospinor} and
${\bf f}_1\cdot{\bf f}_2= ({\bf F}^2-{\bf f}_1^2-{\bf f}_2^2)/2$, 
${\cal P}_2$ and ${\cal P}_4$ can be expressed in terms of 
${\bf f}_1\cdot{\bf f}_2$ and ${\cal P}_0$. 
The resulting expression for the interaction is:
\begin{equation}
{\cal V}({\bf r}_1-{\bf r}_2)=\delta({\bf r}_1-{\bf r}_2)\left
(\alpha +\beta {\bf f}_1\cdot {\bf f}_2+5\gamma {\cal P}_0\right),
\end{equation}
where $\alpha=\frac{1}{7}(4g_2+3g_4)$,  $\beta=-\frac{1}{7}(g_2-g_4)$ and  
$\gamma =\frac{1}{5}(g_0-g_4)-\frac{2}{7}(g_2-g_4)$.
The second-quantized Hamiltonian is then 
\begin{eqnarray}
& & {\cal H}={\cal K} -\int d{\bf r}\  p_0 \psi_a^+ 
(f_{z})_{ab} \psi^{}_b + \nonumber \\
& & \frac{1}{2}\int d{\bf r}(\alpha \psi^{+}_{a}\psi^{+}_{a'}\psi^{}_{a'}
\psi^{}_{a}+  \beta \psi^{+}_{a}\psi^{+}_{a'} 
{\bf f}^{}_{ab}\cdot{\bf f}^{}_{a'b'}\psi^{}_{b'}\psi^{}_{b} \nonumber \\
& & + 5\gamma \psi_a^+\psi_{a'}^+\langle 2a;2a'|00\rangle
\langle 00|2b;2b' \rangle \psi^{}_b \psi^{}_{b'} )
\label{optH}
\end{eqnarray}
where $\langle 00|2b;2b' \rangle$ is the Clebsh-Gordon coefficient for combining
two spin-2 particles with $m_{z} = b$ and $b'$ into a spin singlet
$|0,0\rangle$, 
$\psi_a ({\bf r})$ ($a=2,\ldots -2$) annihilates a Bose at point
${\bf r}$ with spin $a$, and 
${\cal K}= \int {\rm d}{\bf r} \left( \frac{\hbar^{2}}{2M}{\bf
\nabla}\psi^{\dagger}_{a} {\bf \nabla}\psi^{}_{a} + {\cal V}_{\rm trap}
\psi^{\dagger}_{a}\psi^{}_{a} \right)$ is the sum of kinetic energy and 
trap energy ${\cal V}_{\em trap}$. The 
$p_{o}$ term represents the linear Zeeman shift. It
also includes a Lagrange multiplier $p_1$ to constrain the total spin
so that in an external field $B$ along $z$ its expression 
is $p_0=p_1+g\mu_B B\hbar$ \cite{MIT}.
If the ground state $|G\rangle$ is un-fragmented, 
the field operator $\psi_a ({\bf r})$ in eq.(\ref{optH}) becomes a c-number
$\Psi _a({\bf r})=\langle\psi_a({\bf r })\rangle
=\sqrt{n({\bf r})}\zeta_a({\bf r})$, where $n$ is the density 
and the $\zeta$ is a normalized spinor, $\zeta_a^*\zeta^{}_a=1$. 
The ground state energy then becomes 
\begin{equation}
\langle H \rangle_{G} = K -\int np_0\langle f_z\rangle +
\frac{1}{2}\int n^2\left(\alpha+\beta \langle {\bf f}
\rangle ^2 +\gamma |\Theta|^2 \right)\label{H}
\end{equation}
where $K = \langle{\cal K}\rangle_{G}$, 
$\langle {\bf f} \rangle=\zeta^{*T}{\bf f}\zeta$, the 
superscript ``$^T$" stands for 
transpose, and $\Theta =  \sqrt{5} \langle 00|2b; 2b'\rangle \zeta_b
\zeta_{b'}$. More explicitly, $\Theta = \zeta_a 
\hat{A}_{ab}\zeta_b=2\zeta_2\zeta_{-2}-2\zeta_1\zeta_{-1}
+\zeta_0^2$, where $\hat{A}_{ab}= \delta_{a+b,0}(-1)^a$. Note that
$\Theta$ represents a singlet pair of identical spin-2 particles and 
is therefore
invariant under any rotation $U=e^{-i{\bf f}\cdot {\bf c}}$ (where ${\bf c}$
is the rotational angle), i.e. 
$\zeta^{T}\hat{A}\zeta = 
\zeta^{T} U^{T}\hat{A}U \zeta$. This implies 
$\zeta^{T}({\bf f}^{T} \hat{A} + \hat{A}{\bf f}) \zeta =0$ 
for arbitrary $\zeta$, 
which in turn implies that $\hat{A}{\bf f}$ is antisymmetric, and hence 
$\zeta^{T}\hat{A}{\bf f}\zeta=0$. 

It is useful to note that $\Theta$ is the scalar product of a state $\zeta$ with
its time-reversed state $\tilde{\zeta} \equiv \hat{A}\zeta^{\ast}$,
i.e. $\Theta = \tilde{\zeta}^{\dagger}\zeta$.  That $\tilde{\zeta}$ is the
time-reversed state can be seen from the fact that
$\tilde{\zeta}^{\dagger}{\bf f}\tilde{\zeta}$$=$$\tilde{\zeta}^{T}\hat{A}^{T}
{\bf f}\hat{A}\zeta^{\ast}$$=$$- \tilde{\zeta}^{T}{\bf f}^{T} \zeta^{\ast}
= - \zeta^{\dagger}{\bf f}\zeta$. When $\zeta$ is equal to its time-reversed 
partner up to 
a phase factor, (i.e. $\hat{A}\zeta^{\ast} = a\zeta$, $|a|=1$), we 
say that $\zeta$ has no broken time-reversal symmetry. In this case, 
$|\Theta|=1$\cite{TRS}. Time-reversal symmetry
is broken if  $\hat{A}\zeta^{\ast} \neq 
a\zeta$ for any $|a|=1$. 

{\em Determination of Spin Structure}.  The first step to understand the spin
structure is to consider the homogeneous case, where the spin configuration 
$\zeta$ is controlled by the energy function
\begin{equation}
{\cal E}(\zeta)  = \beta \langle {\bf f}\rangle^2
+\gamma |\Theta |^2-p\langle f_z \rangle, \ \ \ \ (p=\frac{2p_0}{n} ).\label{v}
\end{equation}
As we shall see, once the spin structure of the homogenous system 
is determined, the resulting phase diagram will provide a quick though 
qualitative determination of the actual structure. 
It is easy to see from eq.(\ref{v}) that the ground state
magnetization must be aligned with the external field (i.e. along $z$), 
implying $\langle f_+\rangle =0$, where $f_+ = f_{x} + if_{y}$. Eq.(\ref{v})
then becomes 
\begin{equation}
{\cal E} = \beta \langle f_{z} \rangle^2 + \gamma 
|\Theta |^2-p\langle f_z \rangle. 
\label{calE} \end{equation}
Minimizing eq.(\ref{calE}), we have the following Euler-Lagrange equations 
\begin{equation}
(2\beta \langle f_z\rangle-p)\hat{f_z} \zeta-\lambda \zeta 
+2\gamma  \Theta \hat{A}\zeta^*=0, \label{spinoreq}
\end{equation}
where $\lambda$ is the Lagrange multiplier for the 
normalization $\zeta^{\dagger}\zeta =1$. 

By contracting eq.(\ref{spinoreq}) with $\zeta^{\dagger}$ 
(and with $\zeta^T\hat{A}$,
respectively) to the right and using the antisymmetric property of
${\bf f}\hat{A}$, we obtain:
\begin{eqnarray}
& & 2\beta\langle f_z \rangle^2+2\gamma |\Theta|^2 
-p\langle f_z\rangle-\lambda=0
 \label{contract1}\\
& & \Theta(\lambda-2\gamma)=0. \label{contract2}
\end{eqnarray}
Eq.(\ref{contract2}) leads to the following cases: {\bf (a)} $\Theta\neq0$, 
and {\bf (b)} $\Theta=0$. Case {\bf (a)}, which implies $\lambda=2\gamma$, 
 will be referred to as the {\em polar} phase.  In case {\bf (b)}, 
eq.(\ref{spinoreq})-(\ref{contract1}) yield
\begin{equation}
(2\beta\langle f_z\rangle -p)(\hat{f}_z-\langle f_z\rangle)\zeta=0, \label{acases}
\end{equation}
which further divides into : {\bf (b.1)}$(\hat{f}_z-\langle
f_z\rangle)\zeta=0$ ({\em ferromagnetic} phases) and {\bf (b.2)} 
$p=2\beta\langle f_z\rangle$ ({\em cyclic}). 
It is obvious that the ground state is degenerate under gauge transformation 
and spin rotation along $z$. The family of degeneracy is larger when
$p=0$ because of the full rotational symmetry. Without loss of generality, 
we only consider $p>0$.  

{\em Polar phases :}  Setting $\lambda=2\gamma$ in eq.(\ref{spinoreq})
and denoting $\tilde{\zeta}=\hat{A}\zeta^*$, we find
\begin{eqnarray}
& & (2\beta\langle f_z\rangle\hat{f}_z-p\hat{f}_z-2\gamma)\zeta+2\gamma 
\Theta\tilde{\zeta}
=0  \label{realeq1} \\
& & (-2\beta\langle f_z\rangle\hat{f}_z+p\hat{f}_z-2\gamma)\tilde{\zeta}
+2\gamma \Theta^*\zeta
=0 \label{realeq2} \\
& & [(2\beta \langle f_z\rangle -p)^2 {\hat{f}}_z^2-4\gamma^2+4\gamma^2
|\Theta|^2]
\zeta =0 \label{realeq3}
\end{eqnarray}
Eq.(\ref{realeq2}) is obtained by multiplying (\ref{realeq1}) with 
$\hat{A}$
to the right and taking the complex conjugate. Eq. (\ref{realeq3}) shows
$\zeta$ is an eigenstate of $f_{z}^{2}$ with possible 
eigenvalues $f_{z}^{2} = 4, 1, 0$, denoted as ${\bf P}$, ${\bf P1}$, 
and ${\bf P0}$ respectively : 
\begin{eqnarray}
 & {\bf P}:  \, &
\zeta^{T} = \left( 
e^{i\alpha_{2}}\sqrt{\frac{1}{2}+\frac{p}{8\beta-2\gamma}},0,0,0, 
e^{i\alpha_{-2}}\sqrt{\frac{1}{2}-\frac{p}{8\beta-2\gamma}}\right) \nonumber  \\
 & {\bf P1}: \,\, &
\zeta^{T} = \left(0, 
e^{i\alpha_{1}}\sqrt{\frac{1}{2}+\frac{p}{4(\beta-\gamma)}},0,
e^{i\alpha_{-1}}\sqrt{\frac{1}{2}-\frac{p}{4(\beta-\gamma)}}, 0\right)
\nonumber  \\
 & {\bf P0}: \, &
\zeta^{T} = e^{i\alpha_{0}}\left(0,0,1,0,0\right),  
\label{polar} \end{eqnarray}
where 
$\alpha_{i}$ are arbitrary phases.  At $p=0$, $\Theta =1$.

{\em Ferromagnetic phases.} Case {\bf (b.1)} $\Theta =0$ and 
$(\hat{f}_{z} - \langle \hat{f}_{z} \rangle ) \zeta =0$ leads to two 
ferromagnetic phases, 
\begin{equation}
{\bf F} : \,\,\,\,\, \zeta^T=(1,0,0,0,0), \,\,\,\,\,\,\, {\bf F'} : \,\,\,\,\,
\zeta^T=(0,1,0,0,0) 
\end{equation}
with energies ${\cal E}=4\beta-2p$ and ${\cal E} =\beta-p$ respectively.

{\em Cyclic phases}. Case {\bf (b.2)} corresponds to  
$\Theta =0$ and $p=2\beta \langle f_z\rangle $. From eq.(\ref{calE}), 
we see that the states satisfying these two conditions are
degenerate with energy ${\cal E}_{C} = - \frac{p^2}{4\beta}$. Explicitly, 
this degenerate family is specified by the conditions
\begin{eqnarray}
& & \langle f_{+} \rangle \equiv 2\zeta_2^*\zeta_1+\sqrt{6}\zeta_1^*
\zeta_0+\sqrt{6}\zeta_0^*\zeta_{-1}+2\zeta_{-1}^*\zeta_{-2}=0 \\ \label{cyc1}
& & \Theta \equiv 2\zeta_2 \zeta_{-2} -2\zeta_1\zeta_{-1}+\zeta_0^2 =0 \\ 
& & p=2\beta(2|\zeta_2|^2-2|\zeta_{-2}|^2+|\zeta_1|^2-|\zeta_{-1}|^2) \\
& & |\zeta_2|^2+|\zeta_1|^2+|\zeta_0|^2+|\zeta_{-1}|^2+|\zeta_{-2}|^2=1 
\label{cyclast}
\end{eqnarray}
One example of such spinor is 
\begin{equation}
\zeta ^T=\frac{1}{2}(e^{i\phi}(1+\frac{p}{4\beta}),0,
\sqrt{2-\frac{p^2}{8\beta^2}},0,e^{-i\phi}(-1+\frac{p}{4\beta}))
\end{equation}
where $\phi$ is an arbitrary phase.

{\em Phase diagram in zero field.}
Table \ref{t_finite} summarizes the above results. When $p=0$, all polar states
are degenerate.  By searching for the lowest energy
state in Table \ref{t_finite} for given scattering lengths (hence 
given $\beta$ and $\gamma$), we obtain the phase diagram in 
$(\gamma, \beta)$-plane as shown in the inset of Fig.1. 
Among the two ferromagnetic phases only the state ${\bf F}$ is realized. 
${\bf F'}$ is never the lowest energy state for all $(\gamma, \beta)$. 
The three phases ${\bf P}$, ${\bf F}$ and ${\bf C}$ are separated by 
the boundaries 
$\beta=0$ ({\bf F-C}), $\gamma=0$ ({\bf C-P}) and $4\beta=\gamma$ 
({\bf P-F}).

The phases ${\bf P}$, ${\bf C}$, and ${\bf F}$ are characterized by the
``order parameter" 
$(|\Theta|=1, \langle \hat{f}_{z}\rangle =0)$, $(\Theta=0, \langle
\hat{f}_{z}\rangle =0)$, and $(\Theta=0, \langle \hat{f}_{z}\rangle =2)$ 
respectively. Moreover, time-reversal symmetry is broken for the cyclic and the
ferromagnetic states but not for the polar states, where $\zeta$ and
$\hat{A}\zeta^{\ast}$ are related by a phase factor, (see eq.(\ref{polar}))
\cite{TRS}. 
Since the pair $(\Theta, \langle \hat{f}_{z}\rangle)$
undergoes discontinuous changes from one phase to another, the transition
between different phases at $p=0$ are all first order. 

It is also useful to display the phase diagram in terms of the differences 
in scattering lengths $a_0-a_4$ and $a_2-a_4$ as shown in Fig.1. 
The regions occupied by the three phases are : 

\noindent ${\bf P}$ : $a_0-a_4<0$, $\frac{2}{7}|a_2 - a_4|<
\frac{1}{5}|a_0-a_4|$.

\noindent ${\bf F}$ : $(a_2 - a_4)>0$, $\frac{1}{5}(a_0-a_4) + \frac{2}{7}(a_2
- a_4) >0$.

\noindent ${\bf C}$ : $(a_2 - a_4)<0$, $\frac{1}{5}(a_0-a_4) -
\frac{2}{7}(a_2 - a_4) >0$. 

\noindent 
Based on the current estimates of scattering lengths (in a.u.)
by J. Burke and C. Greene\cite{bur},\newline\newline
\begin{tabular}{|c||c|c|c|} \hline
Spin-2 species & $a_0$ & $a_2$ & $a_4$   \\ \hline
$^{23}$Na &  $34.9\pm 1.0$ & $45.8\pm 1.1$  & $64.5\pm 1.3$ \\
$^{87}$Rb &  $89.4\pm 3.0$ & $94.5\pm 3.0$ & $106.0\pm 4.0$ \\
$^{85}$Rb & $-445.0^{+100}_{-300} $ &$-440.0^{+150}_{-225} $ &
$ -420.0^{+100}_{-140}$ \\
$^{83}$Rb & $83.0\pm 3$ & $82.0\pm 3$ & $81.0\pm 3$ \\ \hline
\end{tabular}
\vspace{2mm}\newline
we note from Fig.1 that all the three phases can be realized in the 
above alkali (and in fact in Rb) isotopes. The 
error bars in the estimates of scattering lengths, however, introduce 
uncertainties in the predictions of ground state structures.
The case of $^{87}$Rb is particularly unclear, for it barely resides in 
the polar region while the error bar covers both polar and the 
the cyclic states.  The fact that  all these realizations are 
close to the phase boundaries
means that other physical effects such as gradient energy and magnetic 
field gradients will be important 
in determining the spatial structures of condensate, for they will 
compete with the energy differences between different phases. 

{\em Phase diagram in finite field.} When $p \neq 0$, the phase diagram
depends on the sign of the Heisenberg interaction $\beta$. The phase diagrams 
for $\beta>0$ and $\beta<0$ are shown in Fig.2(a) and 2(b) respectively. 
In all cases, only the polar state ${\bf P}$ is realized as ${\bf P1}$ 
and ${\bf P0}$ are never the lowest energy states. 
When $\beta>0$, (Fig.(2a)), 
the external field $p$ polarizes the 
non-magnetic polar and cyclic states. Since $\Theta$ assumes 
different values in ${\bf P}$ and ${\bf C}$, 
the boundary between them is first order.  As $p$ increases, both 
${\bf P}$ and ${\bf C}$ gradually cross over to the ferromagnetic state. 
The second order phase boundary at which this occurs is 
$p=4\beta$ ($p=4\beta - \gamma$) for the ${\bf C}$ 
(${\bf P}$) state. 
When $\beta<0$, (Fig.(2b)), the only non-magnetic state is
${\bf P}$, which will also be polarized by the external field $p$ 
as in the $\beta >0$ case. The phase boundary between the polar state and the
ferromagnetic state is again a second order line, given by 
$p= -(4|\beta|+\gamma)$. Of course, time-reversal symmetry is broken for all
phases when $p=0$. 

Fig.2(a) and (b) also provide a quick (qualitative) determination of the 
spatial spin structure of non-magnetic phases in optical traps.  
Since $p$ is inversely proportional to the density $n$, (see eq.(\ref{v})),
it generates a vertical trajectory in the phase diagram  as 
one moves from the center of the trap to the surface of the atom cloud. 
Such trajectories 
are shown in Fig(2a) and (2b). The starting points $A$, $B$, and $C$ 
indicate different kinds of spin states at the trap center. The fact that
$p$ diverges at the surface of the cloud means all
non-magnetic states will develop an outer 
ferromagnetic layer whenever $p\neq 0$. 
The above construction does not apply to the ferromagnetic condensate
as its spin structure is not affected by $p$. Numerical calculation including 
the kinetic energy is needed to determine the spatial 
spin configuration with fixed total magnetization. 

Because of the different 
``order parameters" $\Theta$ and $\langle
f_{z} \rangle$, these three phases have very different spinor structures. 
By measuring the
density in each spin component, they can be easily distinguished from each
other.

Finally, we note that there is a one-to-one
corresponding between the structure of a spin-2 spinor Bose condensate 
and that of a d-wave 
BCS superfluids. The latter is known to be characterized by an order 
parameter which is 
a traceless $3\times 3$
symmetry matrix $B_{ij}$, ($B=B^{T}$, Tr$B =0$). 
To see this correspondence, we note from the property
of spherical harmonics that the sum
$P({\bf k}) = k^{2}\zeta_m Y_{2m}(\hat{\bf k})$ is a homogenous polynomial 
of degree 2 in ${\bf k}$ ($P({\bf k}) = B_{ij}k_{i}k_{j}$), and satisfies the 
Laplace equation $\nabla_{k}^{2}P({\bf k}) =0$. These conditions guarantee that 
$B$ is a traceless symmetric matrix.  We can therefore 
associate with each spinor a traceless symmetric matrix $B_{ij} = \int 
\frac{{\rm d}\hat{\bf k}}{4\pi} \hat{k}_{i}\hat{k}_{j} \zeta_{m}Y_{2m}
(\hat{\bf k})$. Rewriting the energy in terms of $B$, one finds that 
eq.(\ref{v}) in zero field ($p=0)$ reduces to the free energy of a d-wave 
superfluid. The exact minimization of this problem was solved 
by N.D. Mermin\cite{Mermin}. Our zero 
field results are in agreement with his exact solution. Because of this 
correspondence, we have named the spinor phases in zero field according to the
features of the d-wave solutions\cite{Mermin}. It is also clear 
from the above discussion that the structure of 
a spin-$S$ spinor condensate ($S$ = integer) 
corresponds to that of a singlet BCS
superfluid with orbital angular momentum  $S$.

We would like to thank Eric Cornell and Carl Wieman for discussions on 
their experiments on $^{85}$Rb, and to Jim Burke and Chris Greene for estimates
of scattering lengths. 
This work was supported by NASA grant NAG8-1441 and NSF
grants DMR-9808125 and DMR-9708274.

{\bf Figure Captions}
Fig.1. Phase diagram in zero field. The ferromagnetic, polar, cyclic 
phases are denoted as ${\bf F}$, ${\bf P}$, and ${\bf C}$ resp.
The phase boundaries ${\bf P-F}$ and ${\bf P-C}$ are straight lines with 
slopes $-7/10$ and $7/10$ resp. The boundary ${\bf F-C}$ is $a_2-a_4 =0$. 
All boundaries are first order lines. The locations of various alkali isotopes
on this diagram are the symbols : $\diamond = ^{23}$Na,
$\times =^{87}$Rb, $\triangle=^{85}$Rb, $\bullet =^{83}$Rb. 
Inset: Zero-field phase diagram in the $(\gamma, \beta)$ plane.\newline

Fig.2. Spinor phase diagrams in finite field for {\bf (a)} $\beta>0$ and
{\bf (b)} $\beta<0$. The thick and dashed lines are first and second 
order phase boundaries resp.  For an atom cloud with a non-magnetic state 
(A, B, or C) at the center, as one moves from the center to the
surface, a vertical line is generated in the phase diagram, showing that the 
outer layer is always ferromagnetic. 

\begin{table}
\begin{tabular}{|c|c|c|} 
 & $\zeta ^T$ & ${\cal E}$  \\ \hline
${\bf F}$  & $(1,0,0,0,0)$ & $4\beta-2p$ \\ 
${\bf F'}$  & $(0,1,0,0,0)$ & $\beta-p$ \\ 
${\bf C}$  & $\frac{1}{2}
\left( e^{i\phi}(1+\frac{p}{4\beta}),0,\sqrt{2-\frac{p^2}{8\beta^2}},
0,e^{-i\phi}(-1+\frac{p}{4\beta})\right)$ & $-\frac{p^2}{4\beta}$ \\ 
${\bf P}$  & $\frac{1}{\sqrt{2}} \left (e^{i\alpha_{2}}\sqrt{1+\frac{p}{4\beta-\gamma}},
0,0,0,e^{i\alpha_{-2}}
\sqrt{1-\frac{p}{4\beta-\gamma}} \right)$ & $\gamma-\frac{p^2}{4\beta-
\gamma}$ \\ 
${\bf P1}$  & $\frac{1}{\sqrt{2}} \left(0, e^{i\alpha_{1}}
\sqrt{1+\frac{p}{2(\beta-\gamma )}},0,
e^{i\alpha_{-1}}
\sqrt{ 1-\frac{p}{2(\beta-\gamma ) }},0 \right)$ & $\gamma-\frac{p^2}
{4(\beta-\gamma)}$ \\ 
${\bf P0}$  & $(0,0,1,0,0)$ & $\gamma$ \\ 
\end{tabular}
\caption{Possible phases in finite field}
\label{t_finite}
\end{table}


\begin{references}
\bibitem{TRS} Any state with $|\Theta|=1$ will not have
broken time-reversal symmetry because $|\zeta - \Theta 
\hat{A}\zeta^{\ast}|^2=0$, which implies $\zeta$ and $\hat{A}\zeta^{\ast}$ are
related by a phase factor. Using $\hat{A}{\bf f} = {\bf
f}^{T}\hat{A}$, it is easy to show that $\langle {\bf f}\rangle =0$ for any
 state with $|\Theta|=1$.
\end{references}
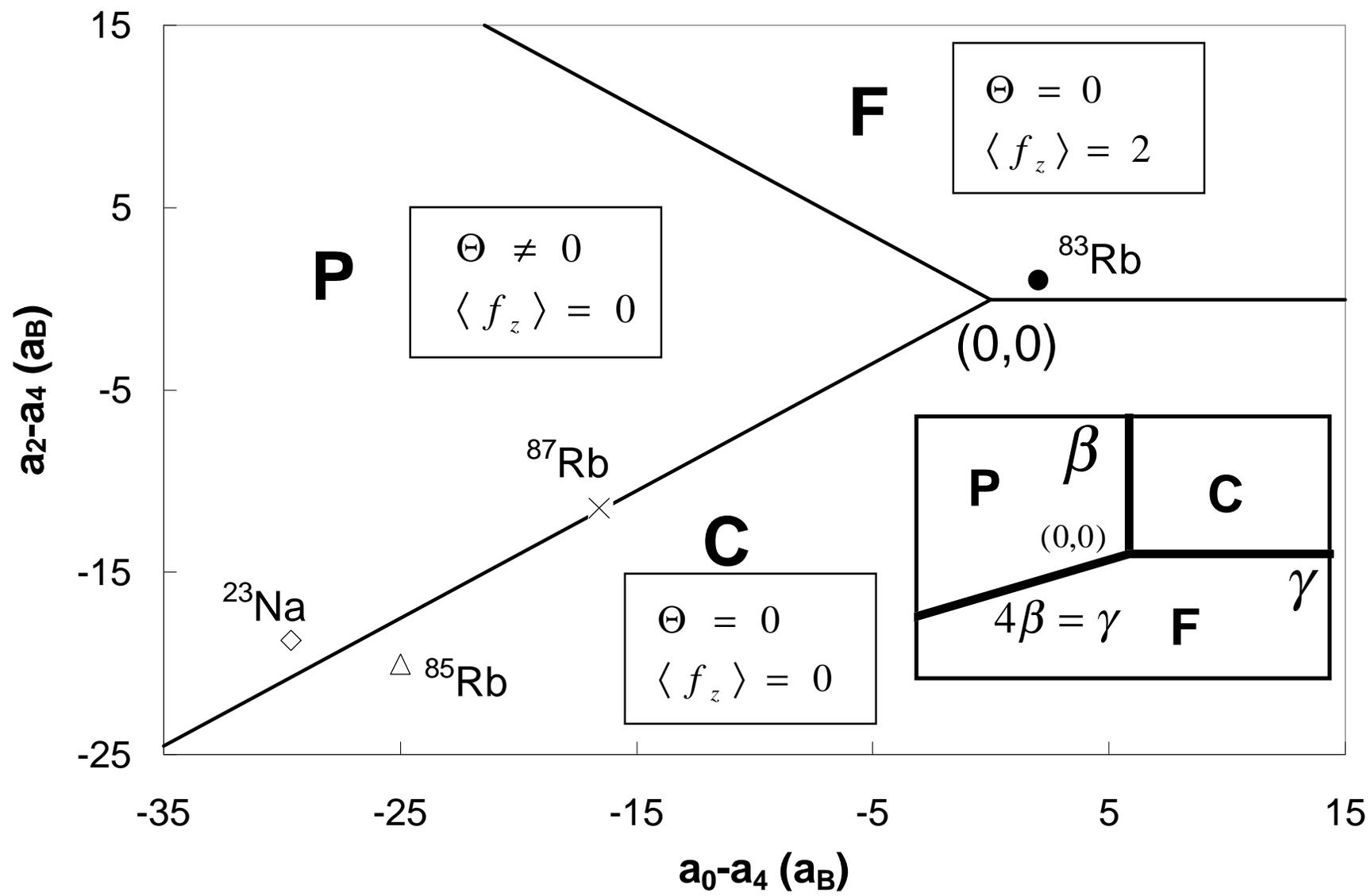
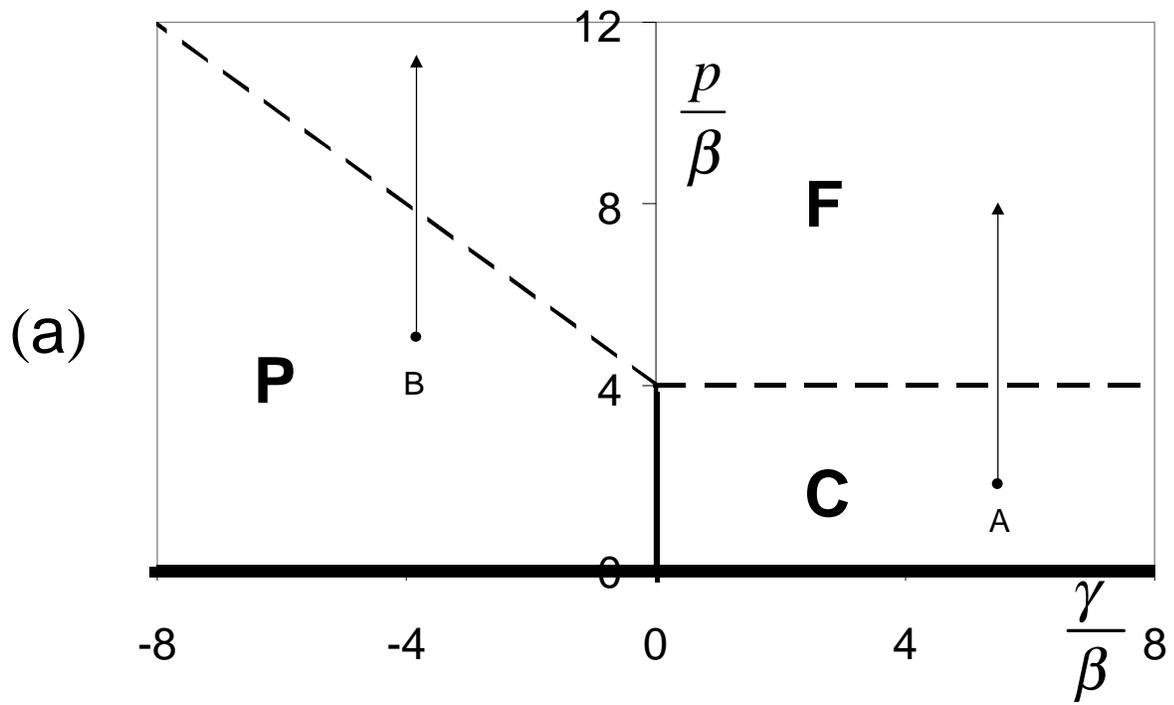
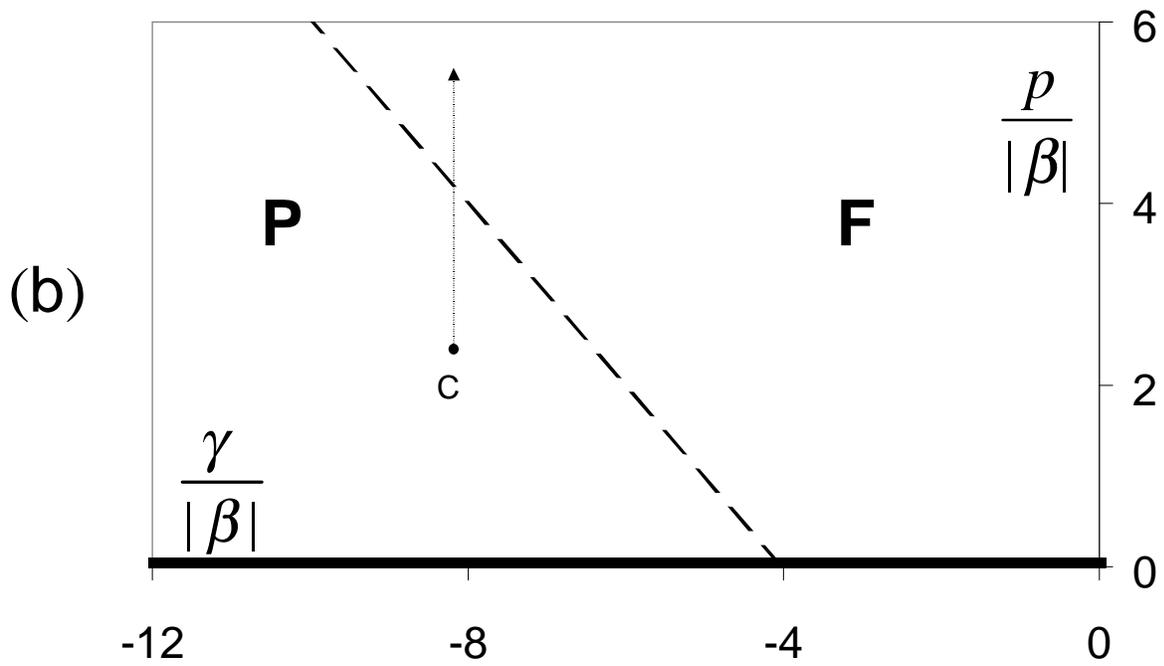
\end{document}